\documentclass[conference]{IEEEtran}
\usepackage{graphicx}
\usepackage{cite}
\usepackage{amsmath,amssymb,amsfonts}
\usepackage{algorithmic}
\usepackage{textcomp}
\usepackage{xcolor}
\usepackage{svg}
\usepackage{xspace}
\usepackage[acronym]{glossaries}
\makeglossaries
\usepackage{multirow}
\usepackage{booktabs} 
\usepackage{url}
\usepackage{color}
\usepackage{multirow}
\usepackage{multicol}
\usepackage{enumitem}
\usepackage{array}
\usepackage[nolist,nohyperlinks]{acronym}
\usepackage{footnote}
\makesavenoteenv{tabular}
\usepackage{breqn}
\usepackage{fancyhdr}
\def\BibTeX{{\rm B\kern-.05em{\sc i\kern-.025em b}\kern-.08em
    T\kern-.1667em\lower.7ex\hbox{E}\kern-.125emX}}

\acrodef{mIoU}{mean Intersection over Union}
\acrodef{DL}{deep learning}
\acrodef{ML}{machine learning}
\acrodef{DSC}{dice coefficient}
\usepackage[acronym]{glossaries}

\acrodef{mIoU}{mean Intersection over Union}
\acrodef{CADx}{Computer-Aided diagnosis}
\acrodef{CNN}{Convolutional Neural Networks}
\acrodef{FPS}{Frame Per Second}
\acrodef{WCE}{Wireless capsule endoscopy}
\acrodef{ASD}{Average Surface Distance}
\acrodef{AI}{Artificial Intelligence}
\acrodef{VCE}{Video capsule endoscopy}
\acrodef{DSC}{Dice Coefficient}
\acrodef{SOTA}{state-of-the-art}
\acrodef{HD}{Hausdorff distance}
\acrodef{MCC}{Matthews correlation coefficient}
\acrodef{GI}{gastrointestinal tract}
\acrodef{OAR}{Organs-at-risk}

\begin{document}
\title{An Efficient Multi-Scale Fusion Network for 3D Organ at Risk (OAR) Segmentation} 
\author{\IEEEauthorblockN{Abhishek Srivastava \IEEEauthorrefmark{1},
Debesh Jha\IEEEauthorrefmark{1},
Elif Keles\IEEEauthorrefmark{1}, 
Bulent Aydogan\IEEEauthorrefmark{2}, 
Mohamed Abazeed\IEEEauthorrefmark{3}, 
Ulas Bagci\IEEEauthorrefmark{1}}

\IEEEauthorblockA{
\IEEEauthorrefmark{1} Machine and Hybrid Intelligence Lab, Department of Radiology, Northwestern University, USA\\ 
\IEEEauthorrefmark{2} Department of Radiation Oncology, University of Chicago, Chicago, IL, USA\\
\IEEEauthorrefmark{3} Department of Radiation Oncology, Northwestern University, Chicago, IL, USA\\
}}

\maketitle
\thispagestyle{fancy}
\begin{abstract}
Accurate segmentation of organs-at-risks (OARs) is a precursor for optimizing radiation therapy planning. Existing deep learning-based multi-scale fusion architectures have demonstrated a tremendous capacity for 2D medical image segmentation. The key to their success is aggregating global context and maintaining high resolution representations. However, when translated into 3D segmentation problems, existing multi-scale fusion architectures might underperform due to their heavy computation overhead and substantial data diet. To address this issue, we propose a new OAR segmentation framework, called \textit{OARFocalFuseNet}, which fuses multi-scale features and employs focal modulation for capturing global-local context across multiple scales. Each resolution stream is enriched with features from different resolution scales, and multi-scale information is aggregated to model diverse contextual ranges. As a result, feature representations are further boosted. The comprehensive comparisons in our experimental setup with OAR segmentation as well as multi-organ segmentation show that our proposed OARFocalFuseNet outperforms the recent state-of-the-art methods on publicly available \textit{OpenKBP datasets} and  \textit{Synapse multi-organ segmentation}. Both of the proposed methods (3D-MSF and {OARFocalFuseNet}) showed promising performance in terms of standard evaluation metrics. Our best performing method (\textit{OARFocalFuseNet}) obtained a dice coefficient of 0.7995 and hausdorff distance of 5.1435 on OpenKBP datasets and dice coefficient 0.8137 on Synapse multi-organ segmentation dataset. 
\end{abstract}


\section{Introduction}
Radiation therapy (RT) is one of the most effective cancer treatments. Approximately half of all cancer patients undergo RT. Maximizing the radiation into the target tumors while minimizing the radiation in non-tumor tissues is the major step in image-based RT applications. This requires delineation of tumor regions as well as identification of organ at risk (OAR) in a precise manner. The exact delineation of all OARs is vital as it prevent from the adverse effects on healthy surrounding organs. Conventionally, expert RT planners manually define OARs from computed tomography (CT) scans, which is tedious and the quality of OARs depends on the expert skills~\cite{Gerhard2021OrganAR}. Recent advances in \ac{DL} have made significant strides in natural and medical image segmentation, however, OAR segmentation remains a challenging task due to heterogeneity of the organ appearance, size, shapes, low contrast of the CT scans, and scanner-related differences~\cite{ibragimov2017segmentation}. Better, more robust, and generalized segmentation algorithms are urgently needed for RT applications.

\ac{CNN}s have served as the defacto architecture for medical image segmentation for the past decade including UNet~\cite{ronneberger2015u} which follows an encoder-decoder based architecture utilizing skip-connections to retain multi-scale features. Since then, extensive research has been done to leverage this design and propose mechanisms which can capture global context. Recent advent of vision transformers has released a wave of methodologies which has further pushed the envelope of performance on computer vision tasks~\cite{dosovitskiy2020image}. Medical image segmentation has also seen an influx of transformer based methods which leverage global context achieved by the self-attention mechanism~\cite{hatamizadeh2022swin,hatamizadeh2022unetr}. Such methods have recently achieved \ac{SOTA} performances on array of medical imaging tasks including the segmentation of multi-organs\cite{chen2021transunet}, cardiac~\cite{chen2020deep}, and polyp~\cite{srivastava2021gmsrf}.

Due to the varying size of the organ of interest for segmentation, multi-scale feature fusion has gained popularity~\cite{gu2022multi,srivastava2021gmsrf,9662196}. However, translation of these models to perform 3D medical image segmentation results in a heavy computation overhead and might result in poor overall performance. To this extent, we design \textit{OARFocalFuseNet} which combines local-global token interactions and multi-resolution feature fusion. We perform experiments on two publicly available datasets with diverse region-of-interest that vary substantially in size and shape. The main contributions of this work are summarized as follows:
\begin{enumerate}
\item We propose a novel \ac{DL} segmentation architecture, OARFocalFuseNet, which upon performing multi-scale fusion utilizes a focal modulation scheme to aggregate multi-scale context in a particular resolution stream. The resultant diverse feature maps can be aided by global-local context and allow OARFocalFuseNet to serve as a benchmark for 3D medical image segmentation.
 \item To ensure a fair comparison with traditional approaches, we also propose a \textit{3D-MSF} (3D-Multi Scale Fusion Network), which fuses multi-scale features in a densely connected mechanism. We use depth-wise convolutions which serve as a parametrically cheaper substitute as compared to its 2D predecessors.
\end{enumerate}

\begin{figure*}[!t]
    \centering
    \includegraphics[width=0.65\textwidth]{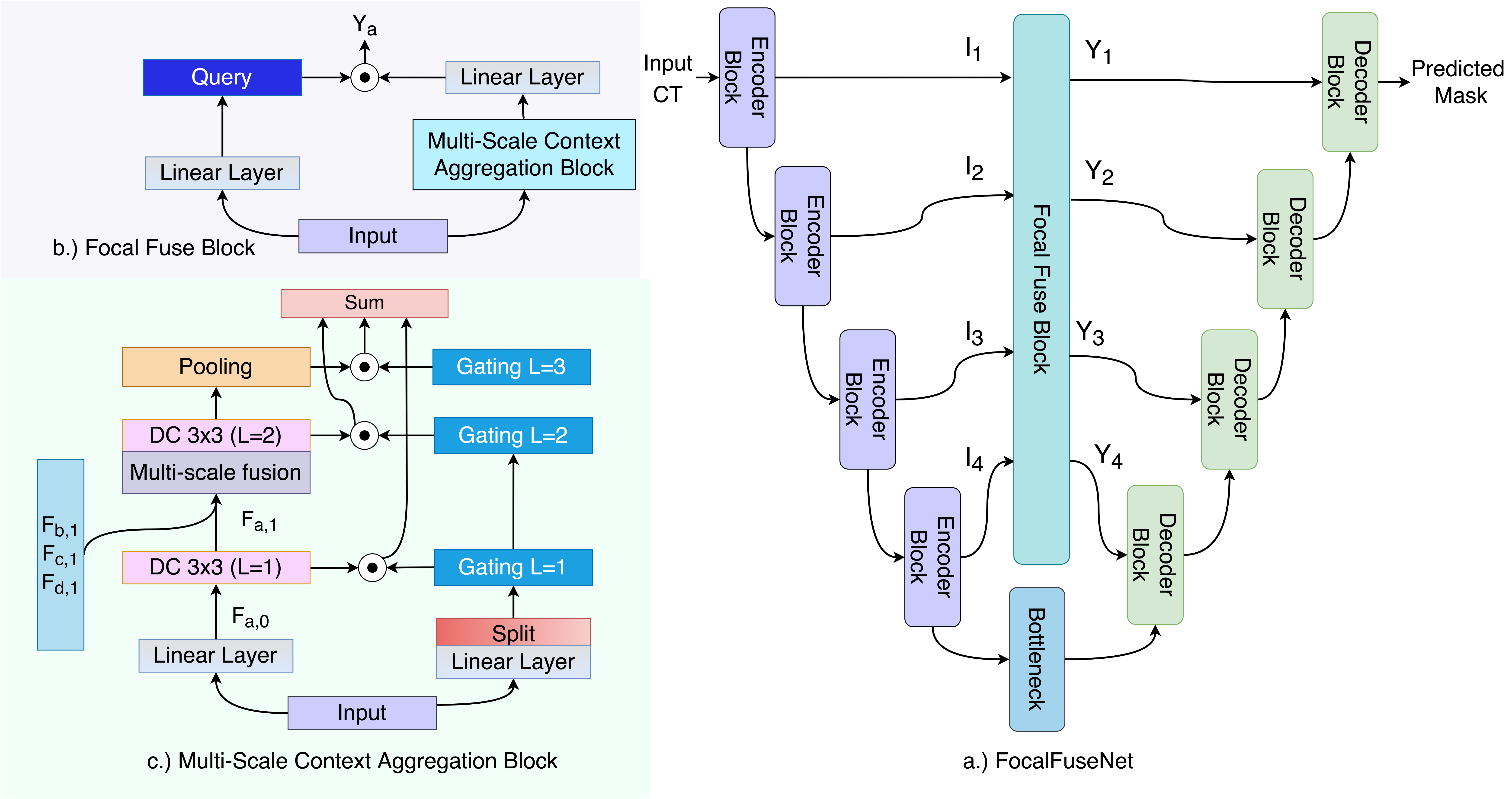}
    \caption{\textbf{The proposed OARFocalFuseNet architectures.} {\bf a}) The complete skeleton of OARFocalFuseNet,  {\bf b}) The Focal Fuse block which aggregates multi-scale global-local context {\bf c}) The Multi-Scale Context Aggregation block which gathers multi-scale features and performs depthwise convolutions to gather features with diverse context ranges and performs spatial and channel wise gating to prune irrelevant features.}
    \label{fig:OARFocalFuseNet}
\end{figure*}

\section{Method}
In this section, we describe the architecture of our two proposed baselines. Section~\ref{subsection:focalfuse} describes the architecture of OARFocalFuseNet and Section~\ref{section:3DMSF} briefly describes the architecture of 3D-MSF.
\subsection{OARFocalFuseNet}
\label{subsection:focalfuse}
\subsubsection{Encoder}
\label{subsubsection:encoder}
Let, $CT$ be the input CT scan where $CT\, \in \, R^{W\:\times\:H\:\times\:Z}$. Here, $W$, $H$, $Z$ are axial, coronal, \& sagittal  axis. Each level/block of the encoder comprises of two successive convolutional layers of kernel size $3$ and a stride of $1$, followed by a pooling layer which halves the dimensions of the feature maps across each axis. We use $4$ consecutive encoder blocks to generate $[I_{1},I_{2},I_{3},I_{4}]$ which are $4$ sets of feature maps with distinct resolutions (see Figure~\ref{fig:OARFocalFuseNet}(a)). An additional encoder block is used as the bottleneck layer to generate $I_{5}$.

\subsubsection{Focal Fuse blocks}
Each of the feature maps $[I_{1},I_{2},I_{3},I_{4}]$ is converted to new feature space using a linear layer(see Equation~\ref{eq:linear} and Figure~\ref{fig:OARFocalFuseNet}(c)).
\begin{equation}
\label{eq:linear}
    F_{a,0} = Linear(I_{a})
\end{equation}
The multi-scale feature fusion is performed by concatenating features from each resolution scale. While propagating feature maps from higher-to-lower resolution streams, we use a single depth-wise convolution operation with a stride of 2 to downsample features, if an additional dimensional reduction is required, a pooling operation is used. Similarly, while transmitting features from lower-to-higher resolution streams, depth-wise transposed convolution operation with stride 2 and if necessary, bicubic interpolation is used. For $N$ focal levels, feature maps generated by the $l$'th multi-scale focal layer are calculated as:
\begin{multline}
    F_{a,l} = GeLU(DC_{3x3}(Conv_{1x1}(F_{a,l-1},F_{b,l-1},F_{c,l-1}, \\
    F_{d,l-1}))) \{b, c, d\} \neq a, \{a, b, c, d\} \in \{1,2,3,4\}
\end{multline}
\label{eq:focal}
Here, $DC$ and $Conv$ represents a depth-wise convolutional layer and a standard convolutional layer, respectively. Additionally, $l$ denotes the multi-scale focal level, which is followed by a $GeLU$ activation layer. Each layer of a particular resolution stream receives input from the previous layers of each resolution stream, which are then fused by a convolutional layer with a stride and kernel size of $1$. The $3\times3$ depthwise convolutional layer incrementally increases the effective receptive field by $3$ after each layer $l$. Hereafter, a global average pooling layer is applied upon the output of the $F_{a,N}$ to obtain the global context. Thus, for each resolution scale $a$ we are able to estimate the local and global context and after each multi-scale focal layer, we communicate information between all the resolution scales (see Figure~\ref{fig:OARFocalFuseNet}(c)). This exchange in features allows each stream to boost the diversity of feature maps by leveraging cross-scale long/short context modelling while maintaining the spatial precision of the feature maps in the high resolution scales. Additionally, we use a linear layer for constraining irrelevant features generated by each multi-scale focal layer (see equation~\ref{eq:gating}).
\begin{equation}
\label{eq:gating}
    G_{a,l} = Linear(F_{a,0}).
\end{equation}
where $G_{a}\, \in \, R^{W\:\times\:H\:\times\:Z\:\times\:N+1}$. The multi-scale focal modulator is generated by adding the context information accumulated by each multi-scale focal layer (see equation~\ref{eq:aggregate} and Figure~\ref{fig:OARFocalFuseNet}(c)).
\begin{equation}
\label{eq:aggregate}
    F_{a} = \sum_{l=1}^{N+1} F_{a,l} \odot  G_{a,l}
\end{equation}
Here, $\odot$ is an element-wise multiplication operator. We use an additional linear layer $F_{a} = Linear(F_{a} \odot G_{a})$ to exchange information amongst channels. The final focally modulated features generated by each scale are calculated as,
\begin{equation}
\label{eq:modeltoken}
    Y_{a} = \sum_{a=1}^{4} F_{a} \odot  Linear(I_{a})
\end{equation}

\begin{table*} [!t]
\centering
\footnotesize
\caption{Comparisons of the results on OpenKBP dataset. We report the mean \ac{DSC}, mean \ac{HD}, mean \ac{ASD}, and \ac{DSC} for each organ.}
\begin{tabular}{@{}l|l|l|l|l|l|l|l|l@{}}
\toprule
\bf{Method} & \bf{Mean DSC} & \bf{Mean \ac{HD}} & \bf{Mean \ac{ASD}} & \bf{Brainstem} & \bf{Spinal cord} & \bf{Right parotid} & \bf{Left parotid} & \bf{Mandible} \\ \hline
UNet-3D~\cite{ronneberger2015u}  & 0.7781 & 5.0662 & 0.6839 & 0.7941 & 0.7444 & 0.7416 & 0.7601 & 0.8503 \\ \hline
AttUNet~\cite{oktay2018attention} & 0.7811 & \bf{4.9981} & 0.6024 & 0.7919 & 0.7439 & 0.7394 & 0.7691 & 0.8611 \\ \hline
DynUNet~\cite{isensee2021nnu} & 0.7931 & 6.3316 & 0.6460 & 0.7958 & 0.7521 & 0.7696 & 0.7731 & 0.8747 \\ \hline
UNETR~\cite{hatamizadeh2022unetr} & 0.7810 & 9.5582 & 0.8527 & 0.7791 & 0.7339 & 0.7610 & 0.7692 & 0.8616 \\ \hline
SwinUNETR~\cite{hatamizadeh2022swin} & 0.7986 & 6.7520 & 0.6409 & \bf{0.8085} & 0.7604 & 0.7706 & 0.7723 & 0.8813 \\ \hline           
\bf{3D-MSF(Ours)}       & 0.7870 & 5.5505 & 0.6152 & 0.7903 & \bf{0.7478} & 0.7498 & 0.7816 & 0.8655 \\ \hline
\bf{OARFocalFuseNet(Ours)} & \bf{0.7995} & 5.1435 & \bf{0.5743} & 0.8031 & 0.7402 & \bf{0.7725} & \bf{0.7987} & \bf{0.8832} \\ \hline 
\bottomrule
\end{tabular}
\label{tab:result1}
\vspace{-5mm}
\end{table*}

\begin{table*}[!t]
\centering
\footnotesize
\caption{Comparisons of results on Synapse multi-organ CT dataset (average dice score and dice score for each organ).}
\begin{tabular}{@{}l|l|l|l|l|l|l|l|l|l|l@{}}
\toprule
\multicolumn{2}{c}{\textbf{Framework}} & \multirow{2}{*}{\bf Mean DSC} & \multirow{2}{*}{\bf Aorta} &  \multirow{2}{*}{\bf Gallbladder} & \multirow{2}{*}{\bf Kidney (L)} & \multirow{2}{*}{\bf Kidney (R)} & \multirow{2}{*}{\bf Liver} & \multirow{2}{*}{\bf Pancreas} & \multirow{2}{*}{\bf Spleen} & \multirow{2}{*}{\bf Stomach} \\
\bf{Encoder} & \bf{Decoder} & & & & & & & &\\ \hline
\multicolumn{2}{c|}{V-Net~\cite{milletari2016v}} &  0.6881 & 0.7534 &  0.5187 & 0.7710 &  0.8075 & 0.8784 & 0.4005 &  0.8056 & 0.5698 \\ \hline
\multicolumn{2}{c|}{DARR~\cite{fu2020domain}} & 0.6977 & 0.7474 &  0.5377 &  0.7231 &  0.7324 & 0.9408 & 0.5418 & 0.8990 & 0.4596 \\ \hline
R50 & U-Net~\cite{ronneberger2015u} & 0.7468 &  0.8418 & 0.6284 & 0.7919 & 0.7129 & 0.9335 & 0.4823 & 0.8441 & 0.7392 \\ \hline
R50 & AttnUNet~\cite{oktay2018attention} & 0.7557 & 0.5592 &  0.6391 &  0.7920 &  0.7271 &  0.9356 & 0.4937 & 0.8719 & 0.7495 \\ \hline
ViT~\cite{dosovitskiy2020image} & None & 0.6150 & 0.4438 & 0.3959 & 0.6746 & 0.6294 &  0.8921 & 0.4314 &  0.7545 &  0.6978 \\ \hline
ViT~\cite{dosovitskiy2020image} & CUP &  0.6786 &  0.7019 & 0.4510 &  0.7470 &  0.6740 & 0.9132 &  0.4200  &  0.8175 &  0.7044 \\ \hline
R50-ViT~\cite{dosovitskiy2020image}&  CUP & 0.7129 &  0.7373 & 0.5513 & 0.7580 & 0.7220 & 0.9151 & 0.4599 &  0.8199 & 0.7395 \\ \hline
\multicolumn{2}{c|}{TransUNet~\cite{chen2021transunet}} &  0.7748 &  0.8723 &  0.6313 &  0.8187 & 0.7702 &  0.9408 & 0.5586 &  0.8508 & 0.7562 \\ \hline
\multicolumn{2}{c|}{3D-MSF(Ours)} & 0.8084 & 0.8883 & \bf{0.6968} & 0.8382 & 0.8204 & 0.9343 & 0.6460 & \bf{0.8705} & \bf{0.7723} \\ \hline
\multicolumn{2}{c|}{\bf OARFocalFuseNet(Ours)} & \bf{0.8137} & \bf{0.9085} & 0.6752 & \bf{0.8424} & \bf{0.8237} & \bf{0.9496} & \bf{0.6808} & 0.8698 & 0.7595 \\ \hline               
\bottomrule
\end{tabular}
\label{tab:result2}
\vspace{-5mm}
\end{table*}

\begin{figure*}[!t]
    \centering
    \includegraphics[width=0.59\textwidth]{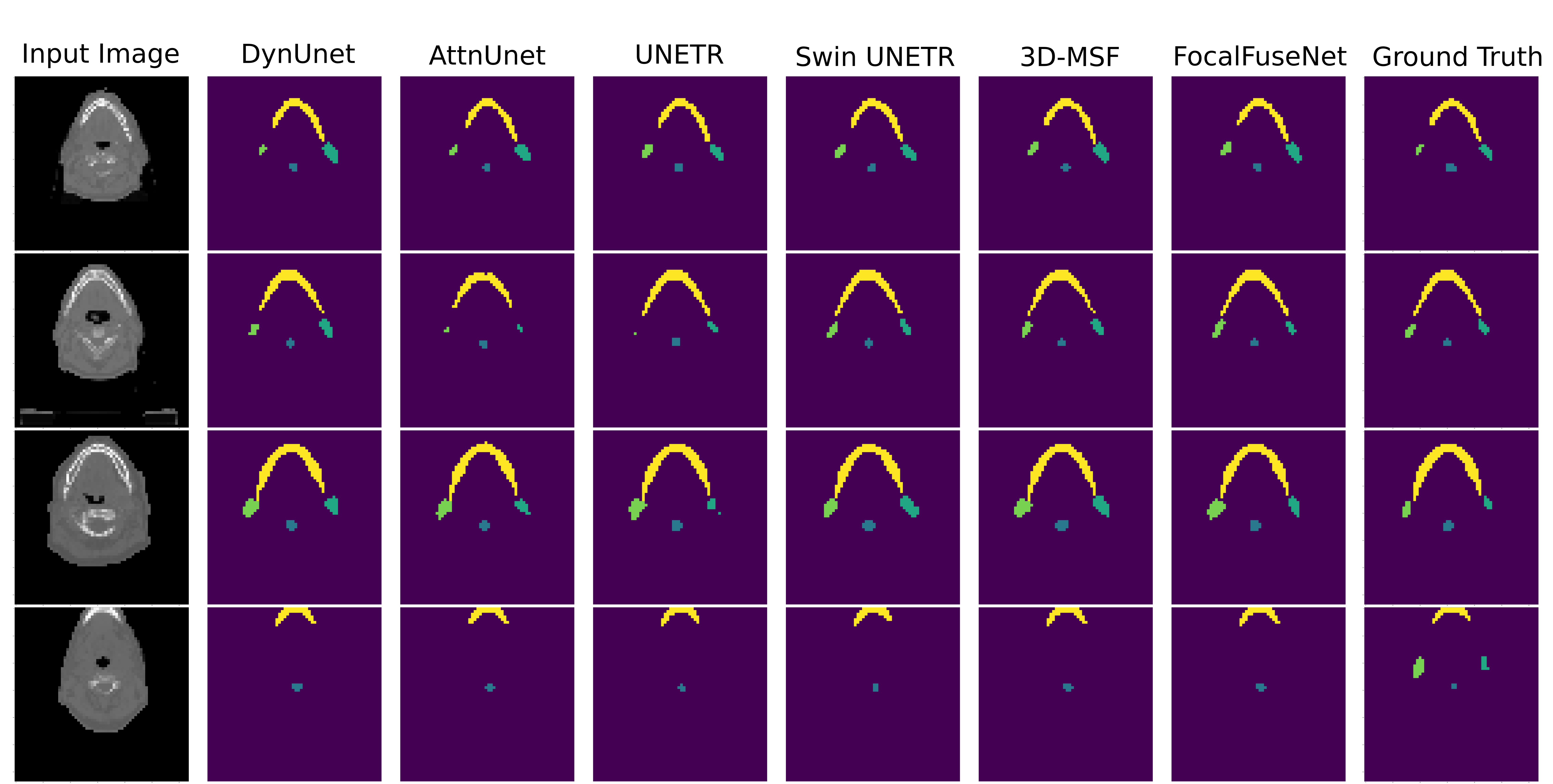}
    \caption{Qualitative comparison of OARFocalFuseNet with other recent benchmarking methods on OpenKBP dataset.}
    \label{fig:qualitative}
    \vspace{-5mm}
\end{figure*}

\subsubsection{Decoder}
\label{subsubsection:decoder}
The first decoder block of OARFocalFuseNet initially receives features from the bottleneck layer and upscales the feature maps by a factor of $2$. Subsequent decoder blocks upscales feature maps as,
\begin{equation}
   D_{x} = ReLU(Norm(Conv_{3\times3}(ConvT(D_{x-1}) \oplus Y_{x})))
   \label{eq:decoder}
\end{equation}
Here, $x \in \{1,2,3,4\}$ and $D_{0}$ is the output of the first decoder block. $ConvT$ is a transposed convolutional layer with kernel size $3$ and stride $2$. $Norm$, $ReLU$ and $\odot$ denotes instance normalization, ReLU activation layer, and concatenation operation, respectively. The final segmentation head uses a convolutional layer which receives the input from $D_{1}$ and has an output number of channels equal to the number of classes. 

\subsection{3D Multi-scale Fusion Network}
\label{section:3DMSF}
The architecture of 3D-MSF consists of the same encoder described in Section~\ref{subsubsection:encoder}. Multi-scale fusion is performed in a densely connected block~\cite{srivastava2021gmsrf}, where inputs to each layer inside a dense block in a particular resolution stream acquires feature maps from all preceding layers within the same stream and the last layer from other resolution streams. For each resolution scale, multi-resolution features are fused as described in equation~\ref{eq:1}.
\begin{equation}
\begin{split}
    I_{i,l} = DepthConv_{3\times3}(I_{a,0} \cdots I_{a,l-1}  \oplus I_{b,l-1} 
    \oplus I_{c,l-1} \\\oplus I_{d,l-1}) , \{b, c, d\} \neq a, \{a,b,c,d\} \in \{1,2,3,4\}
    \end{split}
    \label{eq:1}
\end{equation}
Here, $(a,b,c,d)$ denotes the resolution scale and $l$ denotes the layer inside the dense blocks. The upscaling/downscaling to align spatial resolution of received features from different resolution scales is done identically to the upscaling/downscaling recipe used in OARFocalFuseNet. The same decoder described in Section~\ref{subsubsection:decoder} is used for combining upscaled features from lower decoder layers and features propagated by the skip connections.

\section{Experiments}
\subsection{Datasets and Implementation Details}
\vspace{-1mm}
We evaluate our proposed method on two datasets: \textit{OpenKBP}~\cite{babier2021openkbp} and the \textit{Synapse multi-organ segmentation dataset}\footnote{https://www.synapse.org/\#!Synapse:syn3193805/wiki/217789}. OpenKBP contains data from 340 patients who underwent treatment for head-and-neck cancer while the Synapse multi-organ segmentation dataset contains CT scans and corresponding labels for 13 different organs. The second data set was used to demonstrate that our proposed algorithm is generalizable to other segmentation tasks. We follow the training and testing protocol used in~\cite{isler2022enhancing}, where 188 patients were selected. The selected patients had corresponding labels for five OARs namely: brainstem, spinal cord, right parotid, left parotid, and mandible. Experiments on Synapse multi-organ segmentation dataset followed the same training-testing splits as used in~\cite{chen2021transunet} and evaluated the results on eight classes, namely: aorta, gallbladder, spleen, left kidney, right kidney, liver, pancreas, spleen, and stomach. We set the number of output channels for the first encoder block as $16$ for both OARFocalFuseNet and 3D-MSF. Subsequent encoder blocks increased the number of output channels by a factor of $2$. Adam optimizer was used for optimization. We used a cyclic learning rate scheduler where the maximum learning rate and minimum learning rate were set to $0.003$ and $0.0005$, respectively. We use an equally weighted combination of dice loss and cross-entropy loss for optimization.

\section{Results and Discussion}
\vspace{-1mm}
We provide comparison of our proposed method against other \ac{SOTA} methodologies on Open-KBP~\cite{babier2021openkbp} dataset in Table~\ref{tab:result1}. We can observe that our proposed OARFocalFuseNet has achieved the highest \ac{DSC} of 0.7995 which outperforms the previous \ac{SOTA} method, SwinUNETR~\cite{hatamizadeh2022swin}. Additionally, we observed the lowest \ac{ASD} of $0.5143$ and highest organ-wise \ac{DSC} on segmentation of right parotid, left parotid and mandible. The qualitative comparison of our proposed OARFocalFuseNet with other \ac{SOTA} methods is shown in Figure~\ref{fig:qualitative}. Table~\ref{tab:result2} summarizes the performance of our proposed method against other \ac{SOTA} methodologies on the Synapse multi organ segmentation dataset. Apart from obtaining the highest \ac{DSC} of $0.8137$, we achieved the highest organ-wise \ac{DSC} of $0.9085$, $0.8424$, $0.8237$, $0.9496$, $0.6808$ on segmenting aorta, left kidney, right kidney, liver, and pancreas respectively. Even though convolutional-based multi-scale fusion methods have enjoyed tremendous success in 2-D medical image segmentation, our OARFocalFuseNet outperforms 3D-MSF. This can be attributed to the focal fuse blocks which can maintain high resolution representations while exchanging global and local context across all resolution scales, thereby allowing OARFocalFuseNet to outperform 3D-MSF and other \ac{SOTA} methodologies.

\section{Conclusion}
In this work, we propose the OARFocalFuseNet architecture for \ac{OAR} segmentation, which utilizes a novel focal fuse block for fusing multi-scale features used for capturing long and short range context. We validated our proposed architecture, 3D-MSF and OARFocalFuseNet on two multi-class medical image segmentation datasets, where we showed that the proposed OARFocalFuseNet outperforms other \ac{SOTA} methods in identifying OARs in head-and-neck and abdominal CT scans. In future work, we will examine the failure cases and further modify the architecture to address challenging scenarios in head, neck and abdominal cases.

 \subsubsection*{Acknowledgement}
 This project is supported by the NIH funding: R01-CA246704 and R01-CA240639, and Florida Department of Health (FDOH): 20K04. 

\bibliographystyle{IEEEtran}
\bibliography{references} 

\begin{thebibliography}{10}
\providecommand{\url}[1]{#1}
\csname url@samestyle\endcsname
\providecommand{\newblock}{\relax}
\providecommand{\bibinfo}[2]{#2}
\providecommand{\BIBentrySTDinterwordspacing}{\spaceskip=0pt\relax}
\providecommand{\BIBentryALTinterwordstretchfactor}{4}
\providecommand{\BIBentryALTinterwordspacing}{\spaceskip=\fontdimen2\font plus
\BIBentryALTinterwordstretchfactor\fontdimen3\font minus
  \fontdimen4\font\relax}
\providecommand{\BIBforeignlanguage}[2]{{%
\expandafter\ifx\csname l@#1\endcsname\relax
\typeout{** WARNING: IEEEtran.bst: No hyphenation pattern has been}%
\typeout{** loaded for the language `#1'. Using the pattern for}%
\typeout{** the default language instead.}%
\else
\language=\csname l@#1\endcsname
\fi
#2}}
\providecommand{\BIBdecl}{\relax}
\BIBdecl

\bibitem{Gerhard2021OrganAR}
S.~G. Gerhard \emph{et~al.}, ``Organ at risk dose constraints in sabr: A
  systematic review of active clinical trials.'' \emph{Practical radiation
  oncology}, vol. 11 4, pp. e355--e365, 2021.

\bibitem{ibragimov2017segmentation}
B.~Ibragimov and L.~Xing, ``Segmentation of organs-at-risks in head and neck ct
  images using convolutional neural networks,'' \emph{Medical physics},
  vol.~44, no.~2, pp. 547--557, 2017.

\bibitem{ronneberger2015u}
O.~Ronneberger, P.~Fischer, and T.~Brox, ``U-net: Convolutional networks for
  biomedical image segmentation,'' in \emph{Proceedings of the MICCAI}, 2015,
  pp. 234--241.

\bibitem{dosovitskiy2020image}
A.~Dosovitskiy, L.~Beyer, A.~Kolesnikov, D.~Weissenborn, X.~Zhai,
  T.~Unterthiner, M.~Dehghani, M.~Minderer, G.~Heigold, S.~Gelly \emph{et~al.},
  ``An image is worth 16x16 words: Transformers for image recognition at
  scale,'' \emph{arXiv preprint arXiv:2010.11929}, 2020.

\bibitem{hatamizadeh2022swin}
A.~Hatamizadeh, V.~Nath, Y.~Tang, D.~Yang, H.~Roth, and D.~Xu, ``Swin unetr:
  Swin transformers for semantic segmentation of brain tumors in mri images,''
  \emph{arXiv preprint arXiv:2201.01266}, 2022.

\bibitem{hatamizadeh2022unetr}
A.~Hatamizadeh, Y.~Tang, V.~Nath, D.~Yang, A.~Myronenko, B.~Landman, H.~R.
  Roth, and D.~Xu, ``Unetr: Transformers for 3d medical image segmentation,''
  in \emph{Proceedings of the IEEE/CVF Winter Conference on Applications of
  Computer Vision}, 2022, pp. 574--584.

\bibitem{chen2021transunet}
J.~Chen \emph{et~al.}, ``Transunet: Transformers make strong encoders for
  medical image segmentation,'' \emph{arXiv preprint arXiv:2102.04306}, 2021.

\bibitem{chen2020deep}
C.~Chen \emph{et~al.}, ``Deep learning for cardiac image segmentation: a
  review,'' \emph{Frontiers in Cardiovascular Medicine}, vol.~7, p.~25, 2020.

\bibitem{srivastava2021gmsrf}
A.~Srivastava, S.~Chanda, D.~Jha, U.~Pal, and S.~Ali, ``{GMSRF-Net}: {An
  improved generalizability with global multi-scale residual fusion network for
  polyp segmentation},'' in \emph{Proceedings of the International conference
  on pattern recognition}, 2022.

\bibitem{gu2022multi}
J.~Gu \emph{et~al.}, ``Multi-scale high-resolution vision transformer for
  semantic segmentation,'' in \emph{Proceedings of the IEEE/CVF Conference on
  Computer Vision and Pattern Recognition}, 2022, pp. 12\,094--12\,103.

\bibitem{9662196}
A.~Srivastava, D.~Jha, S.~Chanda, U.~Pal, H.~D. Johansen, D.~Johansen, M.~A.
  Riegler, S.~Ali, and P.~Halvorsen, ``{MSRF-Net}: {A Multi-Scale Residual
  Fusion Network for Biomedical Image Segmentation},'' \emph{IEEE Journal of
  Biomedical and Health Informatics}, vol.~26, no.~5, pp. 2252--2263, 2022.

\bibitem{oktay2018attention}
O.~Oktay \emph{et~al.}, ``Attention u-net: Learning where to look for the
  pancreas,'' \emph{arXiv preprint arXiv:1804.03999}, 2018.

\bibitem{isensee2021nnu}
F.~Isensee, P.~F. Jaeger, S.~A. Kohl, J.~Petersen, and K.~H. Maier-Hein,
  ``nnu-net: a self-configuring method for deep learning-based biomedical image
  segmentation,'' \emph{Nature methods}, vol.~18, no.~2, pp. 203--211, 2021.

\bibitem{milletari2016v}
F.~Milletari, N.~Navab, and S.-A. Ahmadi, ``V-net: Fully convolutional neural
  networks for volumetric medical image segmentation,'' in \emph{Proceedings of
  the fourth international conference on 3D vision (3DV)}, 2016, pp. 565--571.

\bibitem{fu2020domain}
S.~Fu, Y.~Lu, Y.~Wang, Y.~Zhou, W.~Shen, E.~Fishman, and A.~Yuille, ``Domain
  adaptive relational reasoning for 3d multi-organ segmentation,'' in
  \emph{Proceedings of the MICCAI}, 2020, pp. 656--666.

\bibitem{babier2021openkbp}
A.~Babier, B.~Zhang, R.~Mahmood, K.~L. Moore, T.~G. Purdie, A.~L. McNiven, and
  T.~C. Chan, ``Openkbp: The open-access knowledge-based planning grand
  challenge and dataset,'' \emph{Medical Physics}, vol.~48, no.~9, pp.
  5549--5561, 2021.

\bibitem{isler2022enhancing}
I.~Isler, C.~Lisle, J.~Rineer, P.~Kelly, D.~Turgut, J.~Ricci, and U.~Bagci,
  ``Enhancing organ at risk segmentation with improved deep neural networks,''
  in \emph{Medical Imaging 2022: Image Processing}, vol. 12032, 2022, pp.
  814--820.

\end{thebibliography}
\end{document}